\documentclass
[twocolumn,aps,prc,amsmath,showpacs,amssymb,floatfix]
{revtex4}
\usepackage{amssymb}

\usepackage{comment}
\usepackage{enumitem}

\usepackage{CJK}                     
\usepackage[dvips]{graphicx}
\usepackage{bm}                      
\usepackage{mathptmx}                
\usepackage{dcolumn}                 
\usepackage{array}
\begin{document}

\title{Internal X-ray plateau in short GRBs: Signature of supramassive fast-rotating quark stars?}
\author{Ang Li$^{1,2}$\footnote{liang@xmu.edu.cn}, Bing Zhang$^{2,3,4}$\footnote{zhang@physics.unlv.edu}, Nai-Bo Zhang$^{5}$, He Gao$^{6}$, Bin Qi$^{5}$, Tong Liu$^{1,2}$}
\affiliation{
$^1$ Department of Astronomy, Xiamen University, Xiamen, Fujian 361005, China\\
$^2$ Department of Physics and Astronomy, University of Nevada Las Vegas, Nevada 89154, USA\\
$^3$ Department of Astronomy, School of Physics, Peking University, Beijing 100871, China \\
$^4$ Kavli Institute of Astronomy and Astrophysics, Peking University, Beijing 100871, China \\
$^5$ Institute of Space Sciences, Shandong University, Weihai 264209, China\\
$^6$ Department of Astronomy, Beijing Normal University, Beijing 100875, China
}
\date{\today}

\begin{abstract}
A supramassive, strongly-magnetized millisecond neutron star (NS) has been proposed to be the candidate central engine of at least some short gamma-ray bursts (SGRBs), based on the ``internal plateau'' commonly observed in the early X-ray afterglow. While a previous analysis shows a qualitative consistency between this suggestion and the Swift SGRB data, the distribution of observed break time $t_b$ is much narrower than the distribution of the collapse time of supramassive NSs for the several NS equations-of-states (EoSs) investigated. In this paper, we study four recently-constructed ``unified'' NS EoSs (BCPM, BSk20, BSk21, Shen), as well as three developed strange quark star (QS) EoSs within the new confinement density-dependent mass (CDDM) model, labelled as CIDDM, CDDM1, CDDM2. All the EoSs chosen here satisfy the recent observational constraints of the two massive pulsars whose masses are precisely measured. We construct sequences of rigidly rotating NS/QS configurations with increasing spinning frequency $f$, from non-rotating ($f = 0$) to the Keplerian frequency ($f = f_{\rm K}$), and provide convenient analytical parametrizations of the results. Assuming that the cosmological NS-NS merger systems have the same mass distribution as the Galactic NS-NS systems, we demonstrate that all except the BCPM NS EoS can reproduce the current $22\%$ supramassive NS/QS fraction constraint as derived from the SGRB data. We simultaneously simulate the observed quantities (the break time $t_b$, the break time luminosity $L_b$ and the total energy in the electromagnetic channel $E_{\rm total}$) of SGRBs, and find that while equally well reproducing other observational constraints, QS EoSs predict a much narrower $t_b$ distribution than that of the NS EoSs, better matching the data. We therefore suggest that the post-merger product of NS-NS mergers might be fast-rotating supramassive QSs rather than NSs.
\end{abstract}


\maketitle
\section{Introduction}

Short gamma-ray bursts (SGRBs) are generally believed to originate from the mergers of two neutron stars (NS-NS)~\cite{nsns} or one NS and one black hole (NS-BH)~\cite{nsbh}. The nature of their central engine however remains unknown. Recent {\em Swift} observations showed extended central engine activities in the early X-ray afterglow phase~\cite{ee}, in particular the so-called ``internal plateau'', characteristic by a nearly flat light curve plateau extending to $\sim$ 300 seconds followed by a rapid $t^{-(8 \sim 9)}$ decay~\cite{plateaus,lv15}. Since it is very difficult for a BH engine to power such a plateau, one attractive interpretation is that NS-NS mergers produce a rapidly-spinning, supramassive NS~\cite{nsoverbh}, with the rapid decay phase signify the epoch when the star collapses to a BH after it spins down due to dipole radiation or gravitational wave (GW) radiation~\cite{rl14,spindown,gao16}. Whether the current NS modelling could reproduce reasonably all three observed quantities [the break time $t_b$ (or the collapse time), the break time luminosity $L_b$ and the total energy in the electromagnetic channel $E_{\rm total}$] is crucially related to the underlying equation of state (EoS) of dense nuclear matter.

Previous studies showed that some EoS could qualitatively satisfy the observational constraints for individual SGRBs~\cite{lasky14} and large samples~\cite{lv15,gao16}. This justifies and also demands further studies on constraining nuclear matter EoSs from SGRB data. Especially, the recent developments of many-body methods in nuclear physics have enabled a unified treatment~\cite{bcpm,bsk,shen} of all parts of the NS (the outer crust, the inner crust and the core). All the NS EoSs applied so far in the SGRB studies~\cite{lasky14}, however, have been obtained by combing two or three EoSs that handle different density regions of the star, respectively. The matching details at the crust-core interface introduce uncertainties on model calculations~\cite{crustenough}. Therefore, it is essential to use unified NS EoSs to properly address the NS central engine problem of SGRB.

\begin{table*}
\tabcolsep 1pt
\caption{NS/QS EoSs investigated in this study. Here $P_{\rm K}$, $I_{\rm K, max}$ are the Keplerian spin limit and the corresponding maximum moment of inertia, respectively; $M_{\rm TOV}$, $R_{\rm eq}$ are the static gravitational maximum mass by integrating the Tolman-Oppenheimer-Volkoff (TOV) equations and the corresponding equatorial radius, respectively; $\alpha, \beta$ are the fitting parameters for $M_{\rm max}$ in Eq.~(1); $A, B, C$ are the fitting parameters for $R_{\rm eq}$ in Eq.~(2); $a, q, k$ are the fitting parameters for $I_{\rm max}$ in Eq.~(3).}
\vspace*{-12pt}
\begin{center}
\def\temptablewidth{0.96\textwidth}
{\rule{\temptablewidth}{0.5pt}}
\begin{tabular*}{\temptablewidth}{@{\extracolsep{\fill}}cc|cccc|cc|ccc|ccc}
   \hline
   & & $P_{\rm K}$ & $I_{\rm K, max}$ & $M_{\rm TOV}$ & $~R_{\rm eq}~$  & $\alpha$ & $\beta$  & $A$ &  $B$ & $C$  & $a$ & $q$ & $k$ \\
   & EoS & (ms) & $(10^{45} {\rm g~cm}^2)$ & $(M_{\odot})$ & (km) & $(P^{-\beta})$  && $(P^{-B})$ && (km)&& (ms) & $(P^{-1})$ \\
   \hline
   & BCPM  & 0.5584 & 2.857 & 1.98 &  9.941  & 0.03859 & -2.651   & 0.7172 & -2.674 & 9.910 & 0.4509 & 0.3877 & 7.334  \\
NS & BSk20 & 0.5391 & 3.503 & 2.17 & 10.17 & 0.03587 & -2.675   & 0.6347 & -2.638 & 10.18  & 0.4714 & 0.4062 & 6.929  \\
   & BSk21 & 0.6021 & 4.368 & 2.28 & 11.08 & 0.04868 & -2.746   & 0.9429 & -2.696 & 11.03 & 0.4838 & 0.3500 & 7.085 \\
   & Shen & 0.7143 & 4.675 & 2.18 & 12.40 & 0.07657 & -2.738  & 1.393 & -3.431 & 12.47 & 0.4102 & 0.5725 & 8.644 \\
   \hline
   & CIDDM & 0.8326 & 8.645 & 2.09 & 12.43 & 0.16146 & -4.932   & 2.583 & -5.223 & 12.75 & 0.4433 & 0.8079 & 80.76  \\
QS & CDDM1 & 0.9960 & 11.67 & 2.21 & 13.99 & 0.39154 & -4.999   & 7.920 & -5.322 & 14.32 & 0.4253 & 0.9608 & 57.94  \\
   & CDDM2 & 1.1249 & 16.34 & 2.45 & 15.76 & 0.74477 & -5.175   & 17.27 & -5.479 & 16.13 & 0.4205 & 1.087 & 55.14 \\
      \hline
\end{tabular*}
      {\rule{\temptablewidth}{0.5pt}}
\end{center}
\end{table*}

On the other hand, the possibility of a bare quark star (QS; made entirely of de-confined $u, d, s$ quark matter)~\cite{qs} to serve as a the central engine of GRBs has also been discussed by various authors in the past~\cite{QS-GRB}. A recent analysis~\cite{QS-merger} also suggests that the conversion of NSs to QSs is crucial for both SGRBs and long GRBs in the two-families scenario of compact stars, since the well known demanding hyperon puzzle~\cite{hpuzzle} might be a challenge for the existence of massive NSs as heavy as the recent two precisely-measured 2-solar-mass pulsars~\cite{2solar}. We therefore include in the present study the intriguing possibility of a QS engine. In particular, it would be interesting to see whether the observed narrow $t_b$ distribution may be accounted for developed QS EoSs, since NS models could not~\cite{rl14,gao16}. Also, a relatively large ellipticity distribution obtained for NSs~\cite{gao16} is worth further investigation, although it might be explained by the distorting of the inferred strong magnetic fields~\cite{lg16}.

Above all, it is possible to test the proposed post-merger supramassive NS/QS SGRB central engine model with unprecedented accuracy. In this paper, we perform such calculations of rotating NSs and QSs up to their mass-shedding (Keplerian) frequency, by solving exactly the widely-tested $rns$ code~\cite{rns}, and confronting these EoSs with the SGRB data.

\section{NS~EoS~model}

\begin{figure}
\vspace{0.3cm}
{\centering
\resizebox*{0.48\textwidth}{0.3\textheight}
{\includegraphics{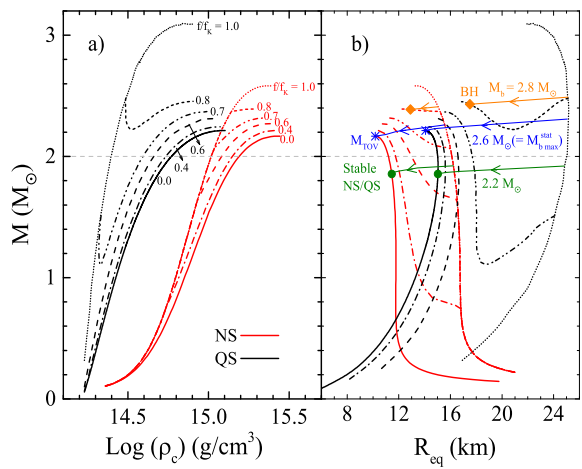}}
\par}
\caption{(Color online) Gravitational mass $M$ vs. central energy density $\rho_{\rm c}$ (panel a) and radius (panel b), for six cases of frequency: $f/f_{\rm K} = 0,~0.4,~0.6,~0.7,~0.8,~1$. Solid lines with arrows denote sequences of constant baryon mass. The NS (QS) results are obtained using the BSk20 (CDDM1) EoS.}\label{fig1}
\end{figure}

The employed unified NS EoSs (BCPM\cite{bcpm}, BSk20, BSk21\cite{bsk}, Shen\cite{shen}) are derived from various many-body framework and cover approximately the full range of high-density models regarding their stellar properties (collected in the first four rows of Table~I). All unified NS EoSs can describe consistently the overall NS structure, which has been quite a challenge due to the difficulties in incorporating additional interactions of the crustal inhomogeneous phase based on nuclear many-body calculations of the core homogeneous matter.

The BCPM EoS, named after Barcelona-Catania-Paris-Madrid, is based on one of the most advanced microscopic approaches, the Brueckner-Hartree-Fock (BHF) theory~\cite{bhf}. The BSk20 and BSk21 EoSs belong to the BSk family of Skyrme nuclear effective forces derived by the Brussels-Montreal group~\cite{bsk}. The high-density part of the two are adjusted to the results of the variational method and the BHF calculations, respectively. The widely-used Shen EoS~\cite{shen} is based on a phenomenological nuclear relativistic mean field model with TM1 parameter set.

\section{QS~EoS~model}

The possible existence of QSs~\cite{qs} originates from a hypothesis back in early 70's~\cite{qm}, namely strange quark matter could be the absolute ground state of matter at zero pressure and temperature. It has intrigued extensive discussions from its detailed phase structures~\cite{qs-phase} and its relevances to various high-energy transient astronomy, such as GRBs~\cite{QS-GRB}, X-ray bursters~\cite{qs-x}, super-luminous
supernovae~\cite{lsn}, and radio pulsars~\cite{qs-psr}.

Although a first-principle calculation in such systems is unachievable due to the complicated nonlinear and non-perturbative nature of quantum chromodynamics (QCD; see \cite{pqcd-ds} for recent progress in perturbative QCD and powerful modeling of QCD in the perturbative and non-perturbative domain using Dyson-Schwinger equations), a comprehensive set of proposed quark-matter EoS~\cite{qs-eos-cddm,qs-eos} has been proposed lately with the basic QCD spirits built in. In the recent version of the confined-density-dependent-mass (CDDM) model~\cite{qs-eos-cddm}, the quark confinement is
achieved by the density dependence of the quark masses
derived from in-medium chiral condensates, and leading-order perturbative interactions have been included. Such terms become dominant at high densities and can lead to absolutely stable strange quark matter and a massive QS made of such matter as heavy as 2 solar mass~\cite{2solar}. In the present calculation, we employ three typical cases of the CDDM QS EoSs~\cite{qs-eos-cddm} (labelled as CIDDM, CDDM1, CDDM2) instead of the simple MIT model~\cite{mit}. The corresponding static QS properties are shown in the last three rows of Table~I. We mention here that the MIT QS EoS model~\cite{mit} allows a more compact QS with $R_{\rm eq}\sim11.5$ km.

\section{ Rotating~NS/QS~configurations}

For a given EoS, the $rns$ code presents uniformly rotating, axisymmetric configurations of a NS/QS. We show them in Fig.~1 for two representative EoSs (BSk20 for NS in red and CDDM1 for QS in black) from the nonrotating cases ($f = 0$) to the Keplerian frequency case ($f = f_{\rm K}$).

We can see that rotation increases the mass that a star of given central density can support. As a consequence, the static configuration with the baryon mass $M_{\rm b} > M_{\rm b, max}^{\rm stat}$ do not exist (in the two EoSs models shown in Fig.~1, $M_{\rm b, max}^{\rm stat} \sim 2.6 M_{\odot}$). Such sequences are so-called {\em supramassive} stars which exist only by virtue of rotation. Those are the ones we are interested in as their spindown-induced collapse to BHs (orange curves in Fig.~1b) would manifest themselves as the rapid decay in X-ray luminosity at the end of plateau. The star sequences of $M_{\rm b} \leq M_{\rm b, max}^{\rm stat}$ (blue and green curves in Fig.~1b) instead would evolve to its static configurations with the same baryon masses as they spin down. Rotation also increases both the equatorial radius and certainly the moment of inertia.

\begin{figure}
\vspace{0.3cm}
{\centering
\resizebox*{0.48\textwidth}{0.29\textheight}
{\includegraphics{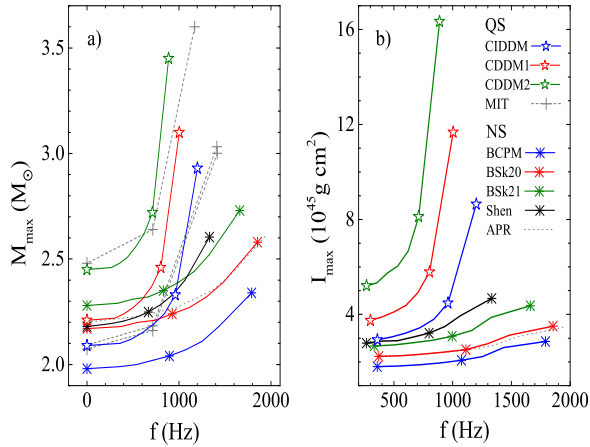}}
\par}
\caption{(Color online) Maximum gravitational mass $M_{\rm max}$ (panel a) and maximum moment of inertia $I_{\rm max}$ (panel b) as a function of the spin frequency, for three cases of QS EoSs (CIDDM, CDDM1, CDDM2) and four cases of unified NS EoSs (BCPM, BSk20, BSk21, Shen). Previous calculations using the APR NS EoS model~\cite{lasky14} and the MIT QS EoS model~\cite{mit} are also shown for comparison. }\label{fig2}
\end{figure}

In Fig.~2 the maximum mass and the maximum moment of inertia are shown as a function of the spin frequency for both NS and QS EoSs. Previous calculations using the APR NS EoS model~\cite{lasky14} and the MIT QS EoS model~\cite{mit} are also shown for comparison. All QS EoS models have similar behaviors but are quite different with various NS EoS models. The $M_{\rm max}$ values for the chosen NS (QS) EoSs are roughly $18-20\%$ ($\sim40\%$) higher than the nonrotating maximum mass $M_{\rm TOV}$. The corresponding increase in $R_{\rm eq}$ is $31-36\%$ ($57-60\%$). Evidently, the increases of ($M_{\rm max}, R_{\rm eq}, I_{\rm max}$) are more pronounced with the QS EoSs than those with the NS ones. We shall soon see that this leads to one main conclusion of the present study that a QS central engine model is more preferred than a NS one.

\begin{figure}[t!]
\vspace{0.3cm}
{\centering
\resizebox*{0.4\textwidth}{0.27\textheight}
{\includegraphics{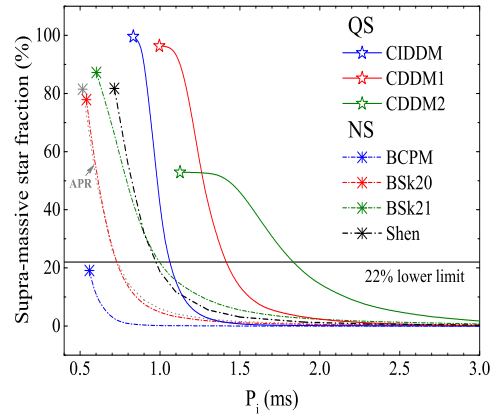}}
\par}
\caption{(Color online) Theoretical estimations of the supramassive star fraction based on four cases of unified NS EoSs and three cases of QS EoSs, as compared with the observed $22\%$ constrain. Previous calculations~\cite{gao16} using the APR NS EoS model are also shown for comparison.}
\label{spectrum}
\end{figure}

\begin{table*}
\tabcolsep 0pt
\caption{Simulated parameter ranges for supramassive NS/QS properties from the {\em Swift} internal plateaus sample with EoS models (except BCPM) from Table I. Here $\epsilon$, $P_i$, $B_p$, $\eta$ are the ellipticity, the initial spin period, the surface dipole magnetic field, and the radiation efficiency, respectively. Data in brackets are those with the best KS tests. $P_{\rm best} (t_b)$ in the last column is the best $P$ value only for the $t_b$ distribution.}
\vspace*{-12pt}
\begin{center}
\def\temptablewidth{0.96\textwidth}
{\rule{\temptablewidth}{0.5pt}}
\begin{tabular*}{\temptablewidth}{@{\extracolsep{\fill}}c|c|c|c|c|c}
\hline
EoS & $\epsilon$ & $P_i~(\rm ms)$ &$ B_p~(G)$ & $\eta$ & $P_{\rm best}~(t_{b})$\\
\hline
BSk20 & $0.002$ &   $0.70-0.75~(0.75)$   &  $N(\mu_{\rm Bp}=10^{14.8-15.4}, \sigma_{\rm Bp}\leq0.2)~[N(\mu_{\rm Bp}=10^{14.9}, \sigma_{\rm Bp}=0.2)]$   & $0.5-1~(0.9)$    & 0.20\\
BSk21 & $0.002$ &   $0.60-0.80~(0.70)$   &  $N(\mu_{\rm Bp}=10^{14.7-15.1}, \sigma_{\rm Bp}\leq0.2)~[N(\mu_{\rm Bp}=10^{15.0}, \sigma_{\rm Bp}=0.2)]$   & $0.7-1~(0.9)$    & 0.29 \\
Shen & $0.002-0.003~(0.002)$ &   $0.70-0.90~(0.70)$   &  $N(\mu_{\rm Bp}=10^{14.6-15.0}, \sigma_{\rm Bp}\leq0.2)~[N(\mu_{\rm Bp}=10^{14.6}, \sigma_{\rm Bp}=0.2)]$   & $0.5-1~(0.9)$    & 0.41 \\
\hline
CIDDM & $0.001$ &   $0.95-1.05~(0.95)$   &  $N(\mu_{\rm Bp}=10^{14.8-15.4}, \sigma_{\rm Bp}\leq0.2)~[N(\mu_{\rm Bp}=10^{15.0}, \sigma_{\rm Bp}=0.2)]$   & $0.5-1(0.5)$    & 0.44\\
CDDM1 & $0.002-0.003~(0.003)$ &   $1.00-1.40~(1.0)$   &  $N(\mu_{\rm Bp}=10^{14.7-15.1}, \sigma_{\rm Bp}\leq0.3)~[N(\mu_{\rm Bp}=10^{14.7}, \sigma_{\rm Bp}=0.2)]$   & $0.5-1(1)$    & 0.65 \\
CDDM2 & $0.004-0.007~(0.005)$ &   $1.10-1.70~(1.3)$   &  $N(\mu_{\rm Bp}=10^{14.8-15.3}, \sigma_{\rm Bp}\leq0.4)~[N(\mu_{\rm Bp}=10^{14.9}, \sigma_{\rm Bp}=0.4)]$   & $0.5-1(1)$    & 0.84\\
\hline
\end{tabular*}
      {\rule{\temptablewidth}{0.5pt}}
\end{center}
\end{table*}

For later use we find that the calculations of $M_{\rm max}, R_{\rm eq}, I_{\rm max}$ can be fitted well as a function of the spin period ($P$) (in millisecond) as follows:
\begin{eqnarray}
&& \frac{M_{\rm max}}{M_{\odot}} = \frac{M_{\rm TOV}}{M_{\odot}} \left[ 1 + \alpha~\left(\frac{P}{\rm ms}\right)^{\beta}\right]; \\
&& \frac{R_{\rm eq}}{\rm km} = C + A~\left(\frac{P}{\rm ms}\right)^B; \\
&& \frac{I_{\rm max}}{10^{45}{\rm g~cm^2}} = \frac{M_{\rm max}}{M_{\odot}} ~ \left(\frac{R_{\rm eq}}{\rm km}\right)^2 \frac{a}{1 + e^{-k~\left(\frac{P}{\rm ms} - q\right)}},
\end{eqnarray}
where the parameters ($\alpha, \beta, A, B, C, a, q, k$) are collected in the last eight columns of Table I.

\section{Supramassive~NS/QS~fraction}

A previous study~\cite{lv15} has identified 21 candidates for supramassive stars (i.e., those bursts with internal plateaus) in 96 SGRBs detected by {\em Swift} up to October 2015. Therefore the current fraction is $\sim 22\%$. Before comparing our results with detailed SGRB observations, it is necessary to first check if the chosen NS/QS EoSs could reproduce such a faction in NS-NS merger products. Such a test is possible if one assumes that the cosmological NS-NS merger systems have the same mass distribution as the Galactic NS-NS binary systems. A distribution of $M = 2.46_{\rm -0.15}^{\rm +0.13} M_{\odot}$ was worked out~\cite{lasky14} for the gravitational mass of the post-merger supramassive stars.

We theoretically calculate, for any given initial spin period $P_i \leq P_{\rm K}$, the upper bound $M_{\rm sup}$ for the mass of the supramassive NS/QS, by solving $[(M_{\rm sup} - M_{\rm TOV})/(\alpha M_{\rm TOV})]^{1/\beta} = P_i$ deduced in the last section. Setting the lower bound as the nonrotating maximum mass $M_{\rm TOV}$, we can finally evaluate the supramassive NS/QS fraction based on the $M = 2.46_{\rm -0.15}^{\rm +0.13} M_{\odot}$ mass distribution~\cite{lasky14}. This is done for all employed NS/QS EoSs. The results are shown in Fig.~3. One can see that all except the BCPM NS EoS can reproduce the $22\%$ fraction constraint (with slightly different required $P_i$). In the following we omit the BCPM EoS.

\section{Collapse~time~simulation~of~supramassive~NSs/QSs}

Previously, when confronting {\em Swift} observations of the internal plateaus sample with several matched NS EoSs~\cite{lasky14}, Ravi \& Lasky~\cite{rl14} and Gao et al.~\cite{gao16} found that although the star parameters can be reasonably constrained, the predicted break time $t_b$ of NSs is always too wide compared with the data. In this section, we apply our previous Monte Carlo simulations \cite{gao16} to the new EoSs for both NSs and QSs studied in this paper. The results are shown in Fig.~4 and Table II.

\begin{figure}
\vspace{0.3cm}
{\centering
\resizebox*{0.4\textwidth}{0.25\textheight}
{\includegraphics{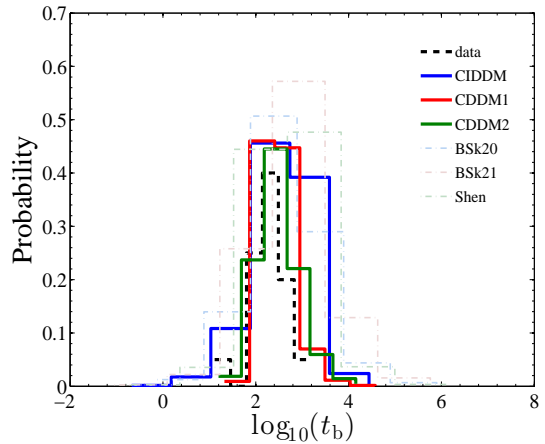}}
\par}
\caption{(Color online) Simulated collapse time distributions with three unified NS EoSs and three QS EoSs, as compared with the observed one (dashed curve).}\label{fig4}
\end{figure}
By requiring that the $P$ values of the Kolmogorov-Smirnov (KS) tests of all three distributions  ($t_b$, $L_b$, $E_{\rm total}$) are larger than $0.05$ as the criteria for reproducing the observations, we list the constrained ranges for the NSs'(QSs') parameters: an ellipticity $\epsilon$ as low as $0.002$ ($0.001$), an initial spin period $P_i$ commonly close to the Keplerian limit $P_{\rm K}$, a surface dipole magnetic field of $B_p \sim 10^{15}$ G, and an efficiency of $\eta = 0.5 - 1$ related to the conversion of the dipole spin-down luminosity to the observed X-ray luminosity. The results with the best $P$ values for the KS tests are listed in brackets. In the last column we show $P_{\rm best} (t_b)$, the best values only for the $t_b$ distribution. It is clear that the KS test for the $t_b$ distribution is indeed improved from Ref.~\cite{gao16}. In particular, as one can see from Fig.~4, the $t_b$ distributions in the QS scenarios are more concentrated, which provide a better agreement with the observed ones. The required $P_i$ for QSs is also larger (longer than 1 ms), which is consistent with the recent numerical simulations of NS-NS mergers that show significant GW is released during the merger phase~\cite{radice16}. Also, a slightly lower and more reasonable magnetic-field-induced ellipticity obtained for QSs are justified by that QSs are more susceptible to magnetic field deformations than NSs~\cite{gjs2012}. We therefore argue that a supramassive QS is favored than a supramassive NS to serve as the central engine of SGRBs with internal plateaus~\footnote{Y.-W. Yu, X.-F. Cao, and X.-P. Zheng, Astrophys. J. \textbf{706}, L221 (2009) reached the similar conclusion based on a completely different argument, i.e. $r$-mode instabilities can be effectively suppressed in QSs.}.

\section{Summary}

To recap, we have carried out the following investigations: 1) Selecting unified NS EoSs that satisfy up-to-date experimental constraints from both nuclear physics and astrophysics, based on modern nuclear many-body theories; 2) Finding typical parameter sets for QS EoSs in developed CDDM model, under same constraints of the NS case for high-density part; 3) Accurately solving the fast-rotating configurations of both NSs and QSs, and providing convenient analytical parameterizations of the results; 4) Checking whether the employed EoSs can fulfill the observed fraction of supramassive stars, based on the mass distribution observation of Galactic NS-NS binary systems; 5) Simulating observed properties of the SGRB internal plateaus sample and revealing the post-merger supramassive stars' physics. We finally reach the conclusion that the post-merger products of NS-NS mergers are probably supramassive QSs rather than NSs. NS-NS mergers are a plausible location for quark de-confinement and the formation of QSs.

\bigskip

\begin{acknowledgments}
This work was supported by the National Basic Research Program (973 Program) of China (No. 2014CB845800), the National Natural Science Foundation of China (No. 11473022, U1331101, U1431107), and the Young Scholars Program of Shandong University, Weihai (Grant No. 2015WHWLJH01).
\end{acknowledgments}



\begin{thebibliography}{99}

\bibitem{nsns}{}
B. Paczynski, Astrophys. J. \textbf{308}, L43 (1986); D. Eichler, M.
Livio, T. Piran, and D.-N. Schramm, Nature (London) \textbf{340},
126 (1989); R. Narayan, B. Paczynski, T. Piran, Astrophys. J. \textbf{395}, L83 (1992); L. Rezzolla, B. Giacomazzo, L. Baiotti, et al. Astrophys. J. \textbf{732}, L6 (2011).

\bibitem{nsbh}{}
B. Paczynski, Acta Astronomica \textbf{41}, 25 (1991).

\bibitem{ee}{}
S.-D. Barthelmy et al., Nature (London) \textbf{438}, 994 (2005);
S. Campana et al., Astron. Astrophys. \textbf{454}, 113 (2006).

\bibitem{lv15}{}
H.-J. L\"{u}, B. Zhang, W.-H. Lei, Ye Li, and P.-D Lasky, Astrophys. J. \textbf{805}, 89 (2015).

\bibitem{plateaus}{}
A. Rowlinson et al., Mon. Not. R. Astron. Soc. \textbf{409}, 531
(2010); Mon. Not. R. Astron. Soc. \textbf{430}, 1061 (2013); E. Troja, et al., Astrophys. J. \textbf{665}, 599 (2007); H.-J. L\"u and B. Zhang, Astrophys. J. \textbf{785}, 74 (2014).

\bibitem{nsoverbh}
Z.-G. Dai, X.-Y. Wang, X.-F. Wu, B. Zhang, Science \textbf{311}, 1127 (2006);
W.-H. Gao, Y.-Z. Fan, Chin. J. Astron. Astrophys. \textbf{6}, 513 (2006);
B.-D. Metzger, A.-L. Piro, E. Quataert, Mon. Not. R. Astron. Soc. \textbf{390}, 781 (2008);
B. Zhang, Astrophys. J. \textbf{763}, L22 (2013);
H. Gao, et al., Astrophys. J. \textbf{771}, 86 (2013).

\bibitem{gao16}{}
H. Gao, B. Zhang, and H.-J. L\"{u}, Phys. Rev. D \textbf{93}, 044065 (2016).

\bibitem{spindown}{}
B. Zhang, Astrophys. J. \textbf{780}, L21 (2014); Y.-Z. Fan, X.-F. Wu, and D.-M. Wei, Phys. Rev. D \textbf{88}, 067304 (2014).

\bibitem{rl14}{}
V. Ravi and P.-D Lasky, Mon. Not. R. Astron. Soc. \textbf{441}, 2433 (2014).

\bibitem{lasky14}{}
P.-D. Lasky, B. Haskell, V. Ravi, E.-J. Howell, and D.-M. Coward, Phys. Rev. D \textbf{89}, 047302 (2014).

\bibitem{bcpm}{}
B.-K. Sharma, M. Centelles, X. Vi\~{n}as, M. Baldo, and G.-F. Burgio, Astron. Astrophys. \textbf{584}, A103 (2015).

\bibitem{bsk}{}
A.-Y. Potekhin, A.-F. Fantina, N. Chamel, J.-M. Pearson, and S. Goriely, Astron. Astrophys. \textbf{560}, A48 (2013).

\bibitem{shen}{}
H. Shen, H. Toki, K. Oyamatsu, and K. Sumiyoshi, Nucl. Phys. A \textbf{637}, 435 (1998).

\bibitem{crustenough}{}
J. Piekarewicz, F.-J. Fattoyev, and C.-J. Horowitz, Phys. Rev.
C \textbf{90}, 015803 (2014).

\bibitem{qs}{}
N.-K. Glendenning, Compact Stars: Nuclear Physics, Particle Physics and General Relativity. Springer, New York (1996); P. Haensel, J.-L. Zdunik, and R. Schaefer, Astron. Astrophys. \textbf{160}, 121 (1986); E. Gourgoulhon, P. Haensel, R. Livine, E. Paluch, S. Bonazzola, and J.-A. Marck, Astron. Astrophys. \textbf{349} 851 (1999); N. Stergioulas, Living Rev. Relat., \textbf{6}, 3 (2003); M. Alford, D. Blaschke, A. Drago, T. Kl\"{a}hn, G. Pagliara, and J. Schaffner-Bielich, Nature \textbf{445}, E7 (2007).

\bibitem{QS-GRB}{}
K.-S. Cheng and Z.-G. Dai, Phys. Rev. Lett. \textbf{77} 1210 (1996); Z.-G. Dai and T. Lu, Phys. Rev. Lett. \textbf{81} 4301 (1998); X.-Y. Wang, Z.-G. Dai, T. Lu, D.-M. Wei, and Y.-F. Huang, Astron. Astrophys. \textbf{357} 543 (2000); R. Ouyed and F. Sannino, Astron. Astrophys. \textbf{387} 725 (2002); A. Drago, A. Lavagno, and G. Pagliara, Phys. Rev. D \textbf{69} 057505 (2004); B. Paczynski and P. Haensel, Mon. Not. R. Astron. Soc. \textbf{362} L4 (2005); R.-X. Xu and E.-W. Liang, Sci. in China G, \textbf{52}, 315 (2009).

\bibitem{QS-merger}{}
A. Drago, A. Lavagno, B. Metzger, G. Pagliara, Phys. Rev. D \textbf{93}, 103001 (2016); A. Drago, A. Lavagno, G. Pagliara, and D. Pigato, Eur. Phys. J. A \textbf{52}, 40 (2016); A. Drago and G. Pagliara, ibid. \textbf{52}, 41 (2016).

\bibitem{hpuzzle}{}
G.-F. Burgio, H.-J. Schulze, and A. Li, Phys. Rev. C \textbf{83}, 025804 (2011); H.-J Schulze and T. Rijken ibid. \textbf{84}, 035801 (2011); Eur. Phys. J. A \textbf{52}, 21 (2016); D. Lonardoni, A. Lovato, S. Gandolfi, and F. Pederiva, Phys. Rev. Lett. \textbf{114}, 092301 (2015).

\bibitem{2solar}{}
P.-B. Demorest, T. Pennucci, S.-M. Ransom, M.-S.-E. Roberts, and J.-W.-T. Hessels, Nature (London) \textbf{467}, 1081 (2010); J. Antoniadis, P.-C.-C. Freire, N. Wex, T.-M. Tauris, R.-S. Lynch et al., Science \textbf{340}, 1233232
(2013).

\bibitem{lg16}{}
P.-D. Lasky and K. Glampedakis, Mon. Not. R. Astron. Soc. \textbf{458}, 1660 (2016).

\bibitem{rns}{}
H. Komatsu, Y. Eriguchi, and I. Hachisu, Mon. Not. R. Astron. Soc. \textbf{237}, 355 (1989); G.-B. Cook, S.-L. Shapiro, and S.-A. Teukolsky, Astrophys. J. \textbf{422}, 227 (1994); N. Stergioulas and J.-L. Friedman, Astrophys. J. \textbf{444}, 306 (1995).

\bibitem{bhf}{}
M. Baldo, \textit{Nuclear Methods and the Nuclear Equation of
State}, International Review of Nuclear Physics Vol. 8 (World
Scientific, Singapore, 1999); W. Zuo, A. Li, Z.-H. Li, and U.
Lombardo, Phys. Rev. C \textbf{70}, 055802 (2004); A. Li, G.-F. Burgio, U. Lombardo, and W. Zuo, ibid. \textbf{74}, 055801 (2006); G.-X. Peng, A. Li, and U. Lombardo, ibid. \textbf{77}, 065807
(2008); A. Li, X.-R. Zhou, G.-F. Burgio, and H.-J. Schulze, ibid. \textbf{81}, 025806 (2010); A. Li, W. Zuo, and G.-X. Peng, ibid. \textbf{91}, 035803 (2015).

\bibitem{qm}{}
A.-R. Bodmer, Phys. Rev. D \textbf{4}, 1601 (1971);
E. Witten, Phys. Rev. D \textbf{30}, 272 (1984).

\bibitem{qs-phase}{}
F. Weber, Prog. Part. Nucl. Phys. \textbf{54}, 193 (2005);
M.-G. Alford, A. Schmitt, K. Rajagopal, and T. Sch$\ddot{a}$fer, 	Rev. Mod. Phys. \textbf{80}, 1455 (2008);
A. Li, R.-X. Xu, and J.-F. Lu, Mon. Not. R. Astron. Soc. \textbf{402}, 2715 (2010).

\bibitem{qs-x}{}
I. Bombaci, Phys. Rev. C \textbf{55} 1587 (1997);
K.-S. Cheng, Z.-G. Dai, D.-M. Wei, and T. Lu, Science \textbf{280}, 407 (1998);
X.-D. Li, S. Ray, J. Dey, M. Dey, and I. Bombaci, Astrophys. J. \textbf{527}, L51 (1999);
X.-D. Li, I. Bombaci, M. Dey, J. Dey, and E.-P.-J. Van Den Heuvel, Phys. Rev. Lett. \textbf{83}, 3776 (1999);
A. Li, G.-X. Peng, and J.-F. Lu, Res. Astron. and Astrophys. \textbf{11}, 482 (2011).

\bibitem{lsn}
Z.-G. Dai, et al., Astrophys. J. \textbf{817}, 132 (2016);
R. Ouyed, D. Leahy, and N. Koning, Astrophys. J. \textbf{818}, 77 (2016).

\bibitem{qs-psr}{}
R.-X. Xu, G.-J. Qiao, and B. Zhang, Astrophys. J. \textbf{522}, L109 (1999);
R.-X. Xu, B. Zhang, and G.-J. Qiao, Astropart. Phys. \textbf{15}, 101 (2001);
R.-X. Xu, Astrophys. J. \textbf{596}, 59 (2003).

\bibitem{pqcd-ds}{}
A. Kurkela, P. Romatschke, and A. Vuorinen, Phys. Rev. D \textbf{81}, 105021 (2010); E. S. Fraga, A. Kurkela, and A. Vuorinen,  Astrophys. J. \textbf{781}, L25 (2014); C. D. Roberts and A. G. Williams, Prog. Part. Nucl. Phys. \textbf{33}, 477 (1994); R. Alkofer, L. von Smekal, Phys. Rep. \textbf{353}, 281 (2001); D. Blaschke, C. D. Roberts, and S. M. Schmidt, Phys. Lett. B \textbf{425}, 232 (1998); C. D. Roberts and S. M. Schmidt, Prog. Part. Nucl. Phys. \textbf{45}, S1 (2000); D. Nickel, R. Alkofer, and J. Wambach, Phys. Rev. D \textbf{74}, 114015 (2006).

\bibitem{qs-eos-cddm}{}
G.-N. Fowler, S. Raha, and R.-M. Weiner, Z. Phys. C. \textbf{9}, 271 (1981); S. Chakrabarty, S. Raha, and B. Sinha, Phys. Lett. B \textbf{229}, 112 (1989); S. Chakrabarty, Phys. Rev. D \textbf{43}, 627 (1991); S. Chakrabarty, ibid. \textbf{48}, 1409 (1993);
S. Chakrabarty, ibid. \textbf{54} 1306 (1996); G.-X. Peng, H.-C. Chiang, B.-S. Zou, P.-Z. Ning, and S.-J. Luo, Phys. Rev. C \textbf{62}, 025801 (2000); C.-J. Xia, G.-X. Peng, S.-W. Chen, Z.-Y. Lu, and J.-F. Xu, Phys. Rev. D \textbf{89}, 105027 (2014); P.-C. Chu and L.-W. Chen, Astrophys. J. \textbf{780}, 135 (2014); A.-I. Qauli and A. Sulaksono, Phys. Rev. D \textbf{93}, 025022 (2016).

\bibitem{qs-eos}{}
M. Dey, I. Bombaci, J. Dey, S. Ray, and B.-C. Samanta, Phys. Lett. B \textbf{438} 123 (1998); M. Buballa, Phys. Rep. \textbf{407}, 205 (2005).

\bibitem{mit}{}
E. Gourgoulhon, P. Haensel, R. Livine, E. Paluch, S. Bonazzola, and J.-A. Marck, Astron. Astrophys, \textbf{349} 851 (1999); S. Bhattacharyya, I. Bombaci, D. Logoteta, and A.-V. Thampan, Mon. Not. R. Astron. Soc. \textbf{457}, 3101 (2016).

\bibitem{radice16}{}
D. Radice, S. Bernuzzi, and C.-D. Ott, arXiv:1603.05726.

\bibitem{gjs2012}{}
K. Glampedakis, D.-I. Jones, and L. Samuelsson, Phys. Rev. Lett. \textbf{109}, 081103 (2012).


\end{thebibliography}
\end{document}